# Ultra-broadband nanophotonic phase shifter based on subwavelength metamaterial waveguides

David González-Andrade,[1,*] José Manuel Luque-González,[2] J. Gonzalo Wangüemert-Pérez,[2] Alejandro Ortega-Moñux,[2] Pavel Cheben,[3] Íñigo Molina-Fernández,[2,4] and Aitor V. Velasco[1]

[1]*Instituto de Óptica Daza de Valdés, Consejo Superior de Investigaciones Científicas (CSIC), Madrid 28006, Spain*
[2]*Departamento de Ingeniería de Comunicaciones, ETSI Telecomunicación, Universidad de Málaga, Málaga 29071, Spain*
[3]*NationalResearch Council Canada, 1200 Montreal Road, Bldg. M50, Ottawa K1A 0R6, Canada*
[4]*Bionand Center for Nanomedicine and Biotechnology, Parque Tecnológico de Andalucía, Málaga 29590, Spain*
*\*Corresponding author: david.gonzalez@csic.es*

**Optical phase shifters are extensively used in integrated optics not only for telecom and datacom applications, but also for sensors and quantum computing. While various active solutions have been demonstrated, progress in passive phase shifters is still lacking. Here, we present a new type of ultra-broadband 90° phase shifter, which exploits the anisotropy and dispersion engineering in subwavelength metamaterial waveguides. Our Floquet-Bloch calculations predict a phase shift error below ±1.7° over an unprecedented operation range from 1.35 $\mu$m to 1.75 $\mu$m, i.e. 400 nm bandwidth covering the E, S, C, L and U telecommunication bands. The flat spectral response of our phase shifter is maintained even in the presence of fabrication errors up to ±20 nm, showing greater robustness than conventional structures. Our device was experimentally demonstrated using standard 220-nm-thick SOI wafers, showing a fourfold reduction in the phase variation compared to conventional phase shifters within the 145 nm wavelength range of our measurement setup. The proposed subwavelength engineered phase shifter paves the way for novel photonic integrated circuits with an ultra-broadband performance.**

## 1. INTRODUCTION

Silicon-on-insulator (SOI) has attracted significant attention in recent years as a promising platform for monolithic integration of optical and electronic circuits [1]. Its compatibility with mature CMOS manufacturing processes has also led to cost-effective and high-volume fabrication of integrated photonic devices such as optical modulators [2], switches [3,4], tunable filters [5,6], and telecom and datacom transceivers [7,8], to name a few. Optical phase shifters (PS) are key components in the aforementioned devices, and in recently proposed large scale quantum silicon photonic circuits [9] for linear quantum computing [10]. Several solutions to achieve phase shifting have been reported in silicon photonics, including active and passive structures, which produce a phase offset between the involved signals by altering the propagation constant of the waveguide modes or by adjusting their optical path lengths.

Active phase shifters are typically narrowband devices that allow to dynamically tune the phase shift response for different wavelengths and also to compensate nominal phase shift deviations arising from fabrication imperfections. This can be achieved by on-chip resistive micro heaters to locally modify the effective refractive index of the waveguides [11], leveraging the high thermo-optic coefficient of silicon [12]. Active phase shifters based on free-carrier plasma dispersion effects and electro-mechanical actuators have also been demonstrated [13-16], with substantially faster response times compared to thermo-optic devices. Active phase shifters present inherent drawbacks such as high power consumption, intricate designs and the need for a control element, which increases the overall device complexity.

Passive phase shifters obviate the requirements for driving power and complex control elements, and are suitable for many applications, including mode-division multiplexing (MDM) [17], 90° hybrids [18], arbitrary-ratio power splitters [19,20], etc., particularly where power consumption is a critical constrain. However, as the response of passive phase shifter cannot be actively tuned, it is also difficult to compensate fabrication errors. Despite the growing interest they have attracted, passive phase shifters have little evolved in the last decade. Most of the structures typically use adiabatic tapers [21] and waveguides with dissimilar lengths [22,23] to modify the optical path, hence induce a phase shift. Phase shifters based on 1×1 multimode interference (MMI) couplers [24] and MMIs with a tilted joint [25] have also been proposed. However, to address the operational requirements for the next generation of photonic integrated circuits, phase shifter's bandwidth and resilience to fabrication errors needs to be significantly improved. Since the early demonstrations of a silicon wire waveguide with a subwavelength grating (SWG) metamaterial core [26,27], metamaterial engineered waveguide structures have emerged as fundamental building blocks for integrated photonics [28,29]. These structures are arrangements of different dielectric materials with a scale substantially smaller than the operating wavelength, hence suppressing diffractive effects. The SWG metamaterials have been successfully used to control refractive index, anisotropy and dispersion in nanophotonic structures [26,30,31,32], including evanescent field sensors [33], spectral filters [34], fiber-to-chip edge couplers [35] and surface grating couplers [36], polarization management devices [37], ultra-broadband directional couplers [38] and MMI devices [39]. For recent comprehensive review see [28,29].

In this work, we propose a new type of 90° phase shifter, leveraging the advantages of SWG metamaterial engineering. The structure is schematically shown in Fig. 1(a). We exploit SWG anisotropy and dispersion engineering to achieve ultra-broadband performance. As a design reference, we also analyze the performance of conventional phase shifters in silicon wire waveguides [see Figs. 1(b) and 1(c)]. Floquet-Bloch simulations of our 90° SWG phase shifter predict a phase deviation below ±1.7° over an unprecedented bandwidth exceeding 400 nm, while conventional phase shifters are limited to ~50 nm. Moreover, fabrication errors up to ±20 nm induce a phase deviation of only 7° over full design wavelength range of our device, compared to 18.7° for conventional PSs. Our experimental results validate simulation predictions, showing a phase slope of only 16°/μm within a 145 nm bandwidth, compared to 64°/μm for the conventional PS structures.

## 2. PRINCIPLE OF OPERATION

Figure 1 shows the schematics of the proposed SWG phase shifter (panel a), as well as two common alternatives known in the state-of-the-art. In all cases, two waveguides of the same length are used to establish a differential phase shift by means of geometric differences in the PS section. Tapered PS shown in Fig. 1(b) comprises two trapezoidal tapers in back-to-back configuration which modify the width of one arm from $W_I$ to $W_{PS}$, while the other arm remains unaltered. Asymmetric PS in Fig. 1(c) utilizes two conventional (continuous) strip waveguides with different widths, $W_U$ and $W_L$, which are connected to the input and output ports via adiabatic tapers. In our SWG PS [Fig. 1(a)], the conventional waveguides are replaced with SWG metamaterial waveguides.

We first investigate bandwidth limitations of conventional phase shifters. We study two parallel Si wire waveguides with different widths, $W_U$ and $W_L$, as shown in the PS section of Fig. 1(c). The accumulated phase difference between both waveguides along the section of length $L_{PS}$ is given by the following expression:

$$\Delta\Phi(\lambda) = \left[\beta_U(\lambda) - \beta_L(\lambda)\right] L_{PS} = \frac{2\pi}{\lambda} \Delta n_{eff}(\lambda) L_{PS} \quad (1)$$

where $\beta_U(\lambda)$ and $\beta_L(\lambda)$ are the propagation constants of the fundamental modes supported by the wide and narrow waveguides, respectively. The free space wavelength is denoted as $\lambda$ and $\Delta n_{eff}(\lambda)$ is the difference between the effective indexes of the fundamental modes propagating through the upper and lower arms, i.e. $\Delta n_{eff}(\lambda) = n_{eff_U}(\lambda) - n_{eff_L}(\lambda)$. The influence of the input and output tapers is considered negligible at this instance due to their comparatively short lengths. Equation (1) shows that the phase shift is primarily governed by the wavelength, given the length $L_{PS}$ is constant. This length, $L_{PS}$, is typically chosen to generate a specific phase shift, $\Delta\Phi_0$, at the design wavelength, $\lambda_0$, according to $L_{PS} = [\lambda_0 \Delta\Phi_0]/[2\pi \Delta n_{eff}(\lambda_0)]$. Thus, the wavelength dependence of the phase shift can be calculated as:

$$\left.\frac{d\Delta\Phi(\lambda)}{d\lambda}\right|_{\lambda=\lambda_0} = -\frac{\Delta\Phi_0}{\lambda_0}\left[1 - \lambda_0 \frac{d\Delta n_{eff}(\lambda)/d\lambda|_{\lambda=\lambda_0}}{\Delta n_{eff}(\lambda_0)}\right]. \quad (2)$$

For comparatively wide waveguides and assuming paraxiality condition holds, propagation constants are $\beta \approx k_0 n_{core} - (\pi\lambda)/(4 n_{core} W_e^2)$ [40] and Eq. (2) can be simplified to $d\Delta\Phi(\lambda)/d\lambda|_{\lambda=\lambda_0} \approx \Delta\Phi_0/\lambda_0$. Here, $k_0$ is the wavenumber, $n_{core}$ is the effective index of the equivalent 2D waveguide and $W_e$ is the effective waveguide width, which is assumed to be invariant with wavelength. This approximation of Eq. (2) unveils the lack of freedom to engineer the dependence on wavelength whereas the choice of greater phase shifts results in narrower bandwidth responses. For example, a 90° phase shifter operating at $\lambda_0 = 1.55\ \mu m$ has a phase slope of $d\Delta\Phi(\lambda)/d\lambda \approx 58°/\mu m$.

Our ultra-broadband phase shifter leverages the inherent anisotropy of subwavelength grating photonic structures. The conventional waveguides are now replaced with two SWG waveguides of widths $W_U$ and $W_L$, both with the same period, $\Lambda$, and duty cycle $DC = a/(a+b)$ [see PS section in Fig. 1(a)]. The SWG waveguides are modelled as a two-dimensional equivalent anisotropic medium described by an effective index tensor: $n_{core} = \text{diag}[n_{xx}, n_{zz}]$ [39], and the accumulated phase shift is:

$$\Delta\Phi_{SWG}(\lambda) \approx \left[\frac{\pi\lambda}{4}\frac{n_{xx}}{n_{zz}^2}\left(\frac{1}{W_{e,L}^2} - \frac{1}{W_{e,U}^2}\right)\right] L_{PS} \quad (3)$$

where $W_{e,L}$ and $W_{e,U}$ are the effective widths of the narrow and wide waveguides, respectively. Halir et al. [39] have recently demonstrated that the term $\lambda n_{xx}/n_{zz}^2$ can be engineered through simulation to mitigate the wavelength dependence inasmuch as the values of $\Lambda$ and DC are judiciously selected. Therefore, the derivative of Eq. (3) is $d\Delta\Phi_{SWG}(\lambda)/d\lambda|_{\lambda=\lambda_0} \approx 0$, and a flat phase shift response is achieved over a broad bandwidth. Hence, SWG metamaterial waveguides offer an interesting opportunity to substantially extend the operational wavelength range of integrated phase shifters.

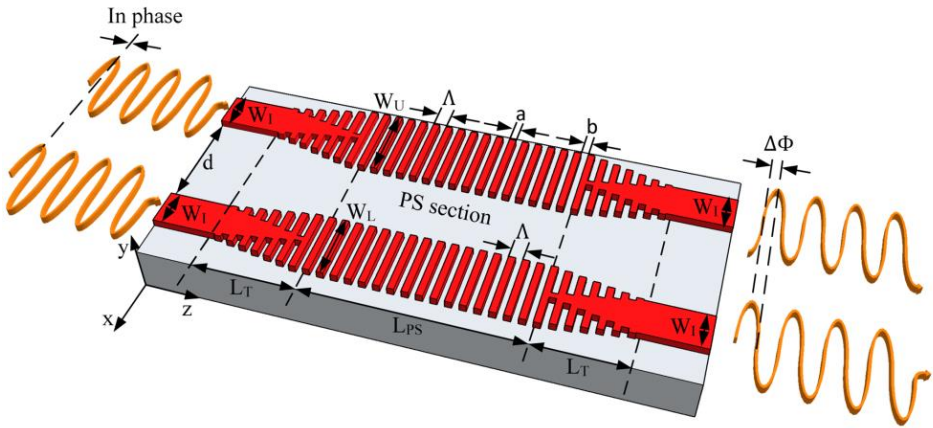
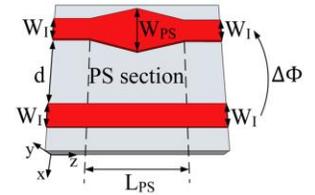
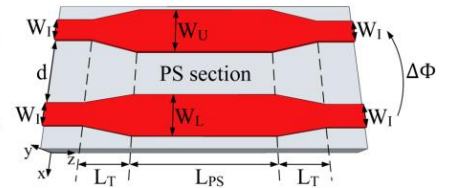

Fig. 1. Schematics of three types of passive phase shifters: (a) our proposed ultra-broadband phase shifter comprising two SWG waveguides with the same period ($\Lambda$) and duty cycle (DC) but with dissimilar widths, $W_U$ and $W_L$; (b) state-of-the-art tapered phase shifter consisting of a straight waveguide and two trapezoidal tapers in back-to-back configuration; and (c) state-of-the-art asymmetric phase shifter based on two non-periodic waveguides with different widths $W_U$ and $W_L$.

## 3. DEVICE DESIGN AND SIMULATION RESULTS

As a reference, we first revisit the performance of conventional phase shifters shown in Figs. 1(b) and 1(c). The separation between the input and output waveguides, $d = 1.5\ \mu m$ is chosen to avoid power coupling, and typical interconnection waveguide widths of $W_I = 0.5\ \mu m$ are assumed. We consider a 220-nm-thick silicon platform surrounded by a silicon dioxide (SiO$_2$) upper cladding and buried oxide (BOX) layer. Si and SiO$_2$ refractive indexes are $n_{Si}(\lambda_0) = 3.476$ and $n_{SiO_2}(\lambda_0) = 1.444$ at the central wavelength of $\lambda_0 = 1.55\ \mu m$. The dispersion of both materials was taken into account in the simulations [41,42].

Tapered PSs can be modelled as the concatenation of multiple sections of parallel waveguides, where the width of one of the arms is different for each section. The length of the PS section is determined by the maximum difference between waveguide widths, i.e. $\Delta W = W_{PS} - 0.5\ \mu m$, and the target phase shift. Modal analysis with Finite Element Method (FEM) [43] was used to design a 90° tapered PS for transverse electric (TE) polarization at $\lambda_0 = 1.55\ \mu m$. The taper width was set to $W_{PS} = 0.7\ \mu m$ and the length was then computed, yielding $L_{PS} = 3.41\ \mu m$. In this design, it is apparent that tapered PSs only possess one degree of freedom, which greatly limits the possibility of reducing wavelength dependence and improving fabrication tolerances. In the asymmetric PS, the width difference between upper and lower arms is selected as $\Delta W = W_U - W_L = 0.2\ \mu m$, for consistency with the previous design. When the widths of both waveguides are wider than 1 $\mu m$, effective indexes of the fundamental modes vary less than for narrow waveguides and the resilience against fabrication errors is improved. For this reason, we choose $W_L = 1.6\ \mu m$ and $W_U = 1.8\ \mu m$. The length of input and output tapers are $L_T = 3\ \mu m$. Taking into account the phase shift introduced by the tapers, the length of the PS section results in $L_{PS} = 36.19\ \mu m$ for a 90° phase shift of the entire structure at $\lambda_0 = 1.55\ \mu m$ and TE polarization.

The figure of merit used to quantify the phase shifter performance is the phase shift error (PSE), which is defined as the deviation from the nominal (90°) phase shift:

$$PSE(\lambda) = \Delta\Phi(\lambda) - 90°. \quad (4)$$

The calculated PSE is shown in Fig. 2 for the designed non-periodic PSs (blue and green curves), yielding almost identical narrowband performance near the central operating wavelength of 1.55 $\mu m$, according to Eq. (2). The response of the tapered PS has a phase slope of ~64°/$\mu m$, whereas the asymmetric PS yields ~71°/$\mu m$, close to the theoretical prediction. Some small differences can be attributed to the use of a 2D model for our first theoretical approximation. Simulations results predict a PSE less than ±13.5° and ±14.4° for tapered and asymmetric PSs, respectively, within the entire simulated wavelength range (1.35 – 1.75 $\mu m$).

To overcome the bandwidth limitations of state-of-the-art phase shifters, we propose to replace conventional waveguides of asymmetric PSs with SWG metamaterial waveguides. As discussed above, we use comparatively wide waveguides, $W_L = 1.6\ \mu m$, in order to increase robustness against fabrication errors. A width difference between the two arms of $\Delta W = 0.2\ \mu m$ is chosen to limit the maximum length of the PS to $L_{PS} \approx 25\ \mu m$ and avoid potential jitter problems in wide SWG waveguides for lengths over 30 $\mu m$ [44]. The dependence with the wavelength is studied by Floquet-Bloch analysis [33] of the SWG waveguides in the PS section. A duty cycle of 50% maximizes the minimum feature size, whereas several SWG period (pitch) values are examined to optimize the bandwidth response [39]. Figure 3(a) shows the PSE as a function of the wavelength for different periods. A very flat response around the central operating wavelength of 1.55 $\mu m$ is found for $\Lambda = 200$ nm, thus ensuring a minimum feature size of 100 nm. The SWG tapers were then designed to perform an adiabatic transition between interconnection and periodic waveguides, yielding a length of $L_T = 3\ \mu m$. The phase shift introduced by the SWG tapers (~20°) was calculated using 3D Finite Difference Time Domain (FDTD) and added to the response of the SWG waveguides shown in Fig. 3(a), by adjusting the length of the PS section. Figure 3(b) shows the PSE as a function of wavelength for $\Lambda = 200$ nm and different number of periods (P), including the influence of SWG tapers. The resolution to adjust the PSE wavelength response is 0.8° per period. We finally select the length $L_{PS} = P \cdot \Lambda = 84 \cdot 0.2 = 16.8\ \mu m$, yielding a minimum PSE over the full simulated bandwidth.

The wavelength response of our SWG phase shifter is also shown in Fig. 2 (red curve), for comparison with conventional devices. It is observed that our subwavelength engineered PS leverages additional degrees of freedom offered by SWG engineering to mitigate the wavelength dependence, achieving an almost flat response with an unprecedented PSE of only ±1.7° over the 400 nm bandwidth.

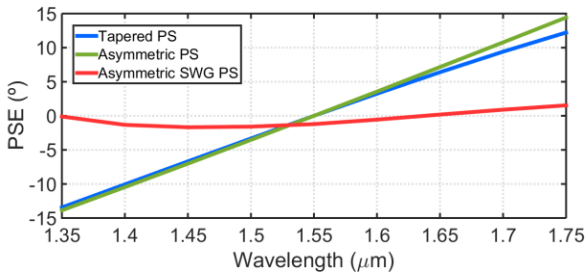

Fig. 2. Comparison of the phase shift error as a function of wavelength for the three designed phase shifters: tapered PS (blue curve), asymmetric PS (green curve) and asymmetric SWG PS (red curve).

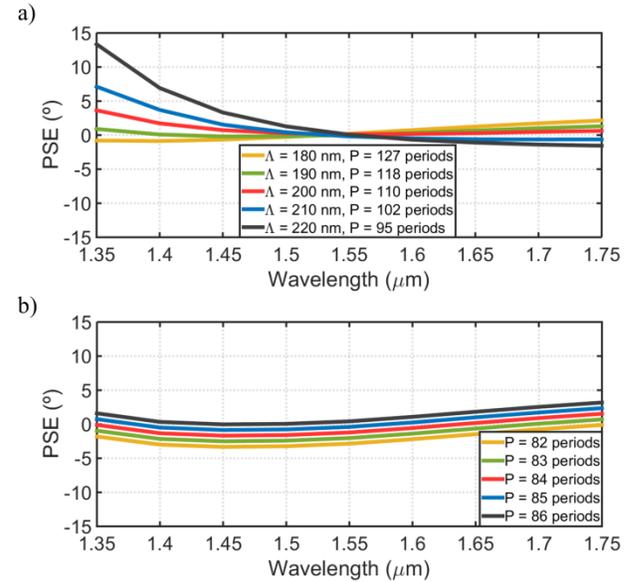

Fig. 3. (a) Phase shift error as a function of wavelength for two parallel SWG waveguides with DC = 50%, $W_L = 1.6\ \mu m$ and $W_U = 1.8\ \mu m$ obtained via Floquet-Bloch analysis. An almost flat response is achieved for $\Lambda = 200$ nm. (b) PSE response of the entire SWG PS with a period $\Lambda = 200$ nm, and including the effect of SWG tapers.

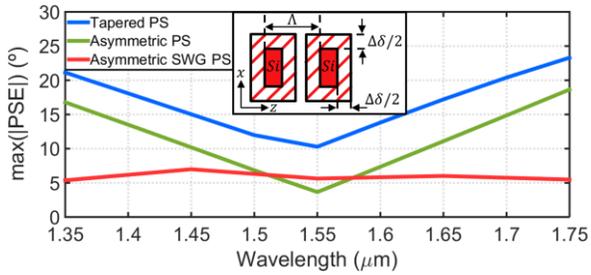

Fig. 4. Simulated maximum phase shift error in the wavelength range 1.35 μm - 1.75 μm and fabrication errors up to $\Delta\delta = \pm 20$ nm. Inset: longitudinal and transversal variations for each SWG segment were considered.

The mesh used for our 3D-FDTD simulations was 20 nm in all directions (transversal and longitudinal), ensuring 10 samples per period (Z axis) and 11 samples along the height of the waveguide (Y axis). Regarding the simulation window, a distance of 0.8 μm was preserved on each side of the waveguides (X axis) and of 1.5 μm on the top and bottom (Y axis). PMLs were used outside the simulation window. Finally, the value of the time step was set as 0.01 μm as proposed by the simulator to meet the condition of Courant. Note that $time\ step = cT$, where $c$ is the speed of light (m/s) and $T$ is the temporary step (s).

Tolerance to fabrication errors was also studied using 3D-FDTD simulations. Dimensional errors of $\Delta\delta = \pm 20$ nm were assumed in both transversal and longitudinal directions for SWG segments [see Fig. 4, inset], thus changing the SWG duty cycle accordingly to the width variations to perform a trustworthy study. For conventional PSs, the influence of waveguide width variation was examined. Figure 4 shows the maximum absolute value of phase shift error for the nominal and biased PSs. It is observed that the worst performing device is tapered PS with a PSE of up to 23.3° within the (1.35 – 1.75 μm) wavelength range. This is due to width changes resulting in an offset of the phase shift curve error, since the length is no longer optimal for the actual PS geometry. This offset is reduced for greater waveguide widths, which leads to a reduced error of 18.7° in the case of asymmetrical PS. In our SWG phase shifter, the maximum phase error is further reduced to 7°, by maintaining the advantage of greater waveguide width of asymmetrical PS, and further benefiting from the reduced effective index inherent to SWG structures. Moreover, the flat spectral response is achieved even in the presence of dimensional errors as large as ±20 nm, yielding a remarkable resilience to typical etching errors.

## 4. FABRICATION AND MEASUREMENTS

Device fabrication was performed in a commercial foundry using a standard SOI wafer with 220-nm-thick silicon layer and 2 μm buried oxide (BOX). The pattern was defined using 100 keV electron beam lithography and a reactive ion etching process with an inductively coupled plasma etcher (ICP-RIE) was used to transfer the pattern to the silicon layer. To protect the devices, a 2.2-μm-thick SiO$_2$ cladding was deposited using a chemical vapour deposition (CVD) process. The experimental characterization of the phase shifters was carried out using a Mach-Zehnder interferometer (MZI) with 14 PSs connected in series, yielding an intensity modulated signal with a period depending on the optical path delay between the arms of the MZI. Owing to space limitations on the chip, only two different types of devices were fabricated, one based on 14 tapered PSs and another based on 14 asymmetric SWG PSs, as shown in Figs. 5(a) and 5(b), respectively. Tapered PSs were included instead of asymmetric PSs, since the former have smaller footprint and similar bandwidth response, thus being typically more used in photonic integrated circuits. In these test structures, the fundamental mode injected to port 1 is equally divided by the SWG engineered MMI, inducing a 90° phase between its outputs. Then, each phase shifter delays the mode propagated through the upper arm an additional 90°, up to a total of 1260° in 14 concatenated phase shifters. Combining both factors and wrapping to the interval [0 – 360°], a phase shift of 270° is achieved, which results in the fundamental mode being coupled into output port 3 of the SWG MMI. When the fundamental mode is injected through port 2, it exists from the output port 4. The accumulated phase errors due to deviations from design central wavelength result in power oscillations between ports 3 and 4 that enables us to characterize the spectral response of each PS.

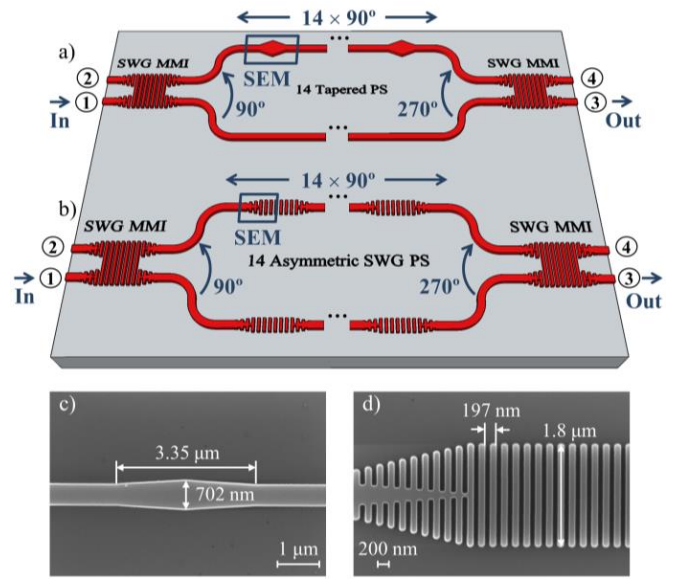

Fig. 5. Schematic of the test structures used to experimentally characterize (a) the tapered PS and (b) the asymmetric SWG PS. Each structure is composed of two ultra-broadband SWG MMIs and 14 phase shifters connected in series, forming a Mach-Zehnder interferometer. Scanning electron microscope images of the fabricated (c) tapered PS and (d) asymmetric SWG PS, as indicated by the blue box in the schematic.

It should be noted that two SWG MMIs were included to ensure ultra-broadband behaviour and circumvent the wavelength limitations of conventional beam splitters in terms of loss, imbalance and phase errors. The dimensions of the SWG MMIs were taken from Halir *et al.* [39], although in our case an optimal length of 77 periods was used for the multimode SWG MMI region. Since modes are more delocalized in SWG waveguides compared to conventional (continuous) waveguides, we increased the separation between the arms of the MZI to 21.5 μm by means of 90° bends, to minimize power coupling. Identical bends were used in the upper and lower arms with a radius of 5 μm, with negligible bend losses for TE polarization [45]. Finally, the dimensions of the phase shifters were taken as specified in the device design section and different flavours varying the number of periods of the PS section were introduced in the mask to compensate for under- or over-etching errors. The best measured performance for the tapered PS was achieved for the nominal design with $L_{PS} \approx 3.41\ \mu m$, while the optimal number of periods for the SWG PS was $P = 86$, i.e. only 2 periods more than the nominal design. Scanning electron microscope images (SEM) of the fabricated PSs are shown in Figs. 5(c) and 5(d).

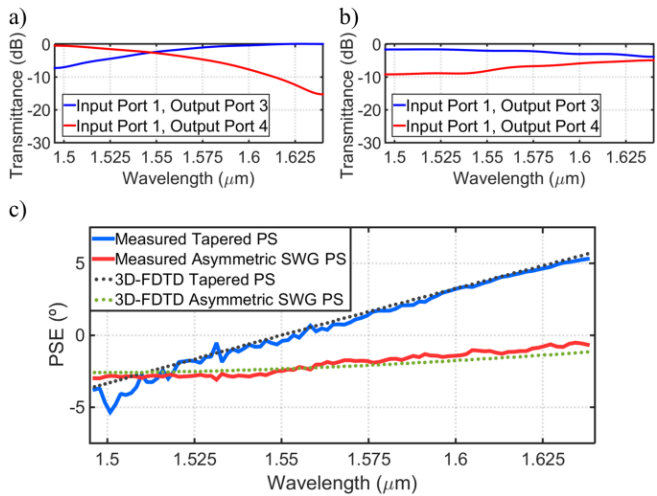

Fig. 6. (a) Measured spectra of the MZIs with 14 tapered phase shifters and (b) with 14 SWG phase shifters. The light was injected through port 1 and both outputs of the test structure (ports 3 and 4) were measured. (c) Measured phase shift error for a single tapered PS (solid blue line) and a single asymmetric SWG PS (solid red line). Dotted lines correspond to the simulation results obtained via 3D-FDTD.

The fabricated devices were characterized using a tunable laser with the wavelength range of $1.495 – 1.64\ \mu m$. The light was coupled in and out of the chip by using high-performance SWG edge couplers [27,35]. Input light polarization was controlled with a lensed polarization maintaining fiber (PMF) assembled in a rotating mount, and TE polarization was selected with a Glan-Thompson polarizer. Transmittance spectra of the MZIs were obtained by sweeping the wavelength of the tunable laser while sequentially measuring the power at both outputs ports with a germanium photodetector placed at the output of the chip.

Negligible insertion losses under 0.2 dB were measured for a single SWG phase shifter. The auxiliary SWG MMI presented losses under 0.6 dB, imbalance below 1 dB and a phase error smaller than 5° within the measured (1.495 – 1.64 μm) wavelength range. Finally, the losses per SWG edge coupler were less than ~4.5 dB. Measured jitter (variations in SWG period) in the SEM images was of the order of only ~3 nm, resulting in negligible impact on the flatness of the spectral response. Furthermore, PS length was maintained under 30 $\mu m$ to avoid potential jitter problems in wide SWG waveguides [44].

The comparison between the measured spectra for the two test structures is shown in Figs. 6(a) and 6(b). The SWG phase shifter shows substantially reduced variations of the output power compared to tapered PSs, within the entire measured wavelength range. It can be observed that maxima and minima from MZI are shifted with respect to the design wavelength of 1.55 $\mu m$. This spectral shift likely originates from the small fabrication errors in each phase shifter, multiplied by a factor of 14 in the overall test structure. To estimate the error introduced by these fabrication defects, we developed a circuit model in which the *S*-parameter matrices of the two SWG MMIs and the 14 PSs were concatenated to obtain the *S*-parameter matrix of the complete MZI. In order to accurately characterize the SWG-MMI and isolate the PSs errors, auxiliary Mach-Zehnder interferometers including SWG-MMIs were fabricated. The resulting experimental losses, imbalance and phase error were used to construct the SWG-MMI *S*-parameter matrix. On the other hand, the phase shifter matrix was constructed with the data of 3D-FDTD simulations, further incorporating a variable offset to characterize the additional phase shift caused by fabrication errors. This offset was then computed by adjusting the curves of the full circuit model to the measured curves through an iterative method. Errors of only -6.5° and 6° were obtained for a single tapered PS and a single asymmetric SWG PS, respectively.

The PSE was derived directly from the transfer functions of the test structures using the measured spectra in Figs. 6(a) and 6(b). For the comparison with simulation results, the phase error introduced by fabrication deviations was subtracted from the measured PSE. Figure 6(c) shows that the measured PSE for both tapered phased shifter (blue curve) and SWG phase shifter (red curve) are in excellent agreement with the 3D-FDTD simulations (dotted lines). The absolute value of the error in the middle of the measured wavelength range is near 2° for both phase shifters, although the measured slope is only 16°/$\mu m$ for our SWG PS, whereas 63°/$\mu m$ is attained for the tapered PS. Note that the phase error response of the SWG PS does not cross zero at the wavelength of 1.55 $\mu m$, because our design was carried out to obtain a minimum error over the simulated wavelength range. The resulting offset in PSE can be compensated by increasing a number of periods in the PS section. Notwithstanding, the concept of flattening the phase response has been experimentally verified for our SWG phase shifter within a measured 145 nm wavelength range, limited by our measurement setup.

## 5. CONCLUSIONS
We have proposed and experimentally demonstrated an ultra-broadband passive phase shifter using subwavelength grating metamaterial structure. Anisotropy and dispersion engineering of SWG waveguides are leveraged to overcome the bandwidth limitations of conventional phase shifters. Our Floquet-Bloch simulations predict an unprecedented phase shift error below ±1.7° within a 400 nm wavelength range (1.35 – 1.75 $\mu m$) for our device, compared to bandwidths of only ~50 nm for conventional devices. Ultra-broadband SWG phase shifters were fabricated on an SOI platform and a very good agreement was found between experimental and simulation results. The phase slope within the measured wavelength range (1.495 – 1.64 $\mu m$) was only 16°/$\mu m$, yielding a fourfold reduction compared to conventional phase shifters. Furthermore, tolerance study shows that SWG devices are more robust to fabrication errors. We believe that SWG phase shifters demonstrated in this paper open promising prospects for the next generation of photonic integrated circuits and could find potential applications in coherent communications, quantum photonics and high-performance mode-division multiplexing circuits for simple and ultra-broadband mode splitters-combiners and higher-order mode converters. Moreover, the application of SWG structures to phase shifters pave the way for future PS applications benefiting from other SWG capabilities such as birefringence engineering for polarization-independent PSs, or waveguide athermalization for temperature-independent PSs.

**Funding.** Spanish Ministry of Science, Innovation and Universities (MICINN) (TEC2015-71127-C2-1-R with FPI scholarship BES-2016-077798, TEC2016-80718-R, IJCI-2016-30484, RTI2018-097957-B-C33); Spanish Ministry of Education, Culture and Sport (MECD)